\documentclass[aps,secnumarabic,notitlepage,twocolumn]{revtex4-1}
\usepackage{amssymb}
\usepackage{amsmath}
\usepackage{graphicx}

\begin{document}

\title{A new view on the superposition of quantum states and the
wave-particle duality of particles}
\author{Yong-Jun Qiao$^{1}$}
\email{qiaodong3601@163.com}
\author{Guo-Feng Zhang$^{2}$}
\email{Corresponding author. gf1978zhang@buaa.edu.cn}
\date{\today }

\begin{abstract}
We construct a coupled quantum vortex superposition state (CVSS), since in
actual physical systems, linear Schr\"{o}dinger equations will not be
available because of a nonlinear effect. By studying the dynamic evolution
of CVSS both analytically and numerically, we show that the superposition of
vortex states is not only a mathematical algebraic sum, but also corresponds
to a physical process of formation. Moreover, a new method to generate
quantum vortex lattice in CVSS research is given. By comparing with the
density profiles and phase distributions of quantum vortex state, we have a
new understanding of vortex state $e^{\pm il\theta}\rightarrow(e^{\pm
i\theta})^{l},L_{z}=\pm l\hbar =\left(\pm \hbar \right) _{1}+\left( \pm
\hbar \right) _{2}+\cdots \left( \pm \hbar \right) _{l}$, which means that
there is spatial degeneracy of angular momentum of a particle. According to
this idea, a free particle can be understood as the center of mass of a
ring-shaped matter in space. Thus, we revisit the double-slit interference
experiment and give a new interpretation.
\end{abstract}

\maketitle

\affiliation{$^{1}$Li Shi senior vocational school, lvliang 033000, China}
\affiliation{$^{2}$School of Physics, Beihang University, Beijing 100191,
China}

\setlength{\parskip}{0.4\baselineskip}

The collapse caused by measuring the quantum superposition state of micro
particles, the non-locality caused by quantum entanglement and the
wave-particle duality reflected by the double-slit interference of particles
are three major problems in the world of quantum mechanics. Schr\"{o}dinger
gave a general equation for describing micro particles, and thus leads to
the emergence of the quantum state superposition hypothesis in 1926 \cite{An
Undulatory Theory of the Mechanics of Atoms and Molecules}. Furthermore, the
application of entanglement theory is already brilliant before its
principles are fully understood \cite{Quantum entanglement,18 Qubit
Entanglement,Experimental quantum teleportation,Entanglement of
photons,Einstein Podolsky Rosen,Entanglement of photons with complex,Can
Quantum Mechanical Description}. And in order to clarify the principle of
the double-slit interference experiment of particles, Bohm proposed the
guided wave theory \cite{A Suggested Interpretation of the Quantum Theory in
Terms of Hidden Variables I and II}, but its fatal flaw is that the guided
wave speed is greater than the speed of light.

Strictly speaking, the linear quantum superposition state only satisfies the
linear Schr\"{o}dinger equation, but the generation mechanism and stability
of the vortex-antivortex superposition state are studied in the
Bose-Einstein condensate (BEC) with non-zero scattering between particles
which cause a nonlinear effect \cite{Structure and generation of the
vortex-antivortex superposed state in Bose-Einstein condensates,The Vortex
Phase Qubit,Structure of two-component Bose-Einstein condensates with
respective}, and experimentally there have been successful cases by way of
Josephson coupling \cite{Optical control of the internal and external
angular momentum of a Bose-Einstein condensate,Quantized Rotation of Atoms
From Photons with Orbital Angular Momentum,Sculpting the Vortex State of a
Spinor BEC,Spontaneous breaking of spatial and spin symmetry in spinor
condensates,Arbitrary Coherent Superpositions of Quantized Vortices}. So we
suggest a coupled vortex superposition state (CVSS) which contains a hidden
variable $\kappa $ (Josephson coupling coefficient). In the study of its
dynamic evolution, we find that the collapse caused by quantum measurement
is no longer so dramatic, and the non-local entangled system is localized
into a vortex space. Furthermore, the double-slit experiment of particles
shows that the particle source can emit particles with a intrinsic angular
momentum (IAM) of $\frac{\hbar }{2}\cos 2\kappa t$ or $-\frac{\hbar }{2}\cos
2\kappa t$ which determining whether the interference fringes appear or not
on screen.

In this letter, we investigate the dynamic evolution of CVSS both
analytically and numerically. It is shown that the superposition of vortex
states is not only a mathematical algebraic sum, but also corresponds to a
physical process of formation.\emph{\ }Meanwhile,\emph{\ }we discover a new
method of generating quantum vortex lattice in BEC,\ and get a new
understanding of the quantum vortex state, namely $e^{\pm il\theta
}\rightarrow \left( e^{\pm i\theta }\right) ^{l}$ and angular momentum $%
L_{z}=\pm l\hbar =\left( \pm \hbar \right) _{1}+\left( \pm \hbar \right)
_{2}+\cdots \left( \pm \hbar \right) _{l}$\textbf{. }That is to say, the
vortex angular momentum of a particle has degenerate nature of spatial
position. According to this idea, we find that the particles in entangled
state in CVSS have an IAM of $\frac{\hbar }{2}\cos 2\kappa t$ or $-\frac{%
\hbar }{2}\cos 2\kappa t$, so there is a hidden variable $\kappa $ in the
entangled state. In addition, we also show that a free particle have an IAM
of $\hbar /2$ or $-\hbar /2$, and it can be understood as the center of mass
of a ring-shaped matter in space, we design a scheme and give the numerical
simulation of the double-slit interference of particles as a solid proof.

\textbf{Coupled vortex superposition state---}We suggest there is a linear
vortex superposition state (LVSS) $\psi =\frac{\sqrt{2}}{2}({\psi
_{l_{1}}+\psi _{l_{2}})}$ to particles, which satisfy the linear Schr\"{o}%
dinger equation. However, the preparation of quantum superposition state by
Josephson coupling is a common method for researchers, so we make a new
analysis of dynamics evolution of LVSS as shown in Fig. 1. Here vortices are
${\psi _{l_{j}}(r,\theta ,t)}=f_{j}(r)e^{il_{j}\theta }e^{-i\mu _{j}t}$ with
$f_{j}(r)=A_{j}e^{-r^{2}/2\sigma _{j}^{2}}(r/\sigma _{j})^{|l_{j}|}$, and $%
A_{j}^{2}=\frac{1}{\pi \sigma _{j}^{2}{|l}_{j}{|}!}$ $(j=1,2)$ being a
normalization constant as $\iint \left\vert \psi \right\vert ^{2}rdrd\theta
=1$, $l_{j}$ is the angular momentum quantum number of a vortex, $\sigma
_{j} $ and $\mu _{j}$ are the size of the particles distribution scale in
space and the chemical potential of particles, respectively.
\begin{figure}[th]
\centering\includegraphics[width=6cm,height=3cm]{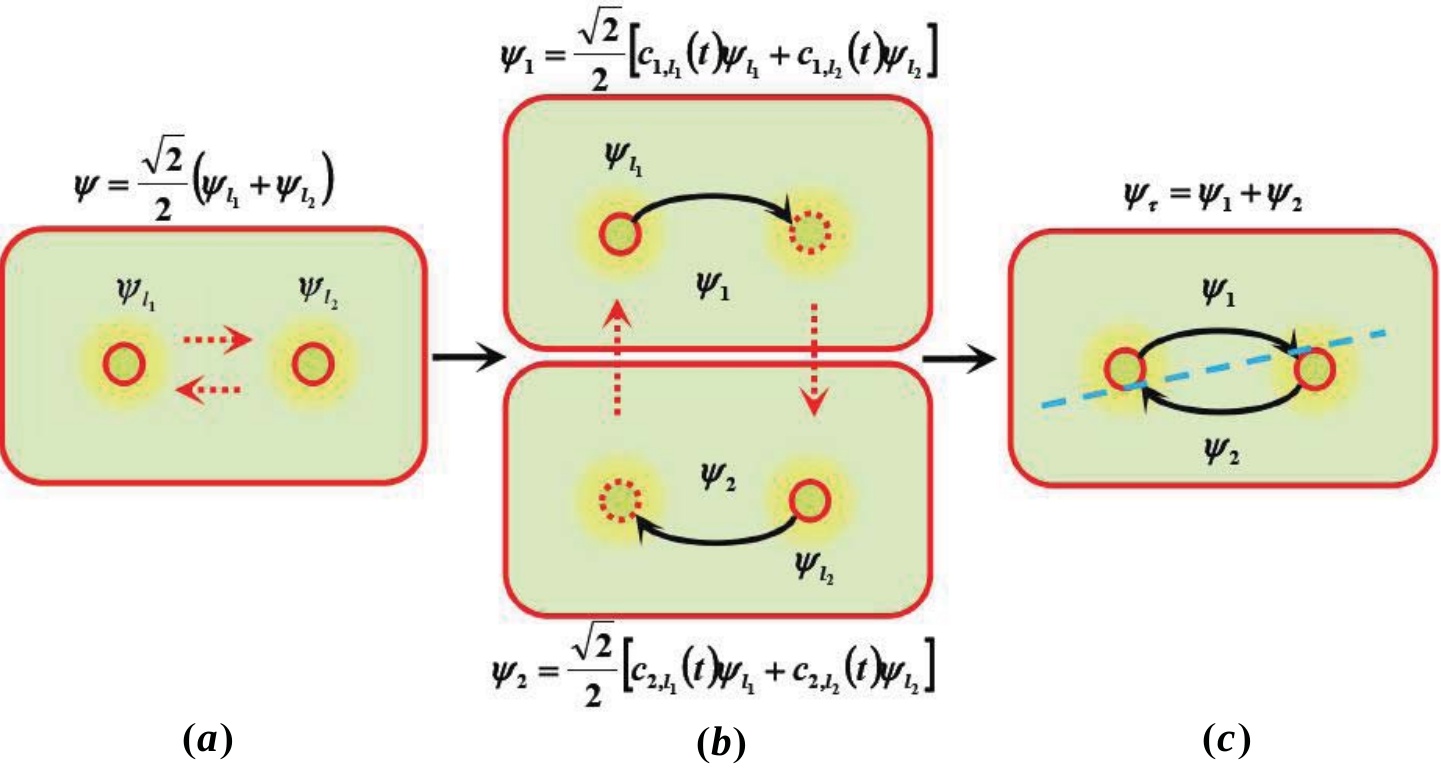}
\caption{Schematic diagram of LVSS which is defined as CVSS. (a) Quantum
vortex superposition state is prepared experimentally by Josephson coupling,
and in theory it is expressed as the algebraic sum of the quantum vortex
states. (b) Quantum states $\protect\psi _{1}$ and $\protect\psi _{2}$
formed by $\protect\psi _{l_{1}}$ and $\protect\psi _{l_{2}}$ after
Josephson coupling. (c) Coupled vortex superposition state generated by $%
\protect\psi _{1}$ and $\protect\psi _{2}$. The $\protect\sqrt{2}/2$ in $%
\protect\psi _{1}$ and $\protect\psi _{2}$ show the sequential transition
between them for particles.}
\end{figure}
Now we assume $\psi _{1}$ and $\psi _{2}$ satisfy Eq. (1) with $\iint
\left\vert c_{j,l_{1}}(t){\psi _{l_{1}}+c_{j,l_{2}}(t)\psi _{l_{2}}}%
\right\vert ^{2}rdrd\theta =1$, $\iint \left\vert \psi _{\tau }\right\vert
^{2}rdrd\theta =1$. And $\psi _{\tau }$\ is the actual superposition state
named the coupled vortex superposition state (CVSS)\textbf{. }

\begin{eqnarray}
i{\frac{\partial \psi _{1}}{\partial t}} &=&[-\bigtriangledown _{\perp
}^{2}+V(r)+g_{1}|\psi _{1}|^{2}+g_{12}|\psi _{2}|^{2}]\psi _{1}  \notag \\
&&-\kappa \psi _{2},  \notag \\
i{\frac{\partial \psi _{2}}{\partial t}} &=&[-\bigtriangledown _{\perp
}^{2}+V(r)+g_{2}|\psi _{2}|^{2}+g_{12}|\psi _{1}|^{2}]\psi _{2}  \notag \\
&&-\kappa \psi _{1}.
\end{eqnarray}%
Here $\nabla _{\bot }^{2}=\frac{1}{r}\frac{\partial }{\partial r}(r\frac{%
\partial }{\partial r})+\frac{1}{r^{2}}\frac{\partial ^{2}}{\partial \theta
^{2}}$ is the 2D operator in a $x-y$ plane that we have adopted plane-polar
coordinates $(r,\theta )$, and $V(r)=V_{0}r^{2}$ is a trapping potential.
Here $g_{1}$, $g_{2}$, $g_{12}$\ are non-linear constants which characterize
the scattering interaction between particles. Note that $\psi _{1}$ and $%
\psi _{2}$ are two states of particles which are in one-component partcle
system, rather than describing a two-component\textbf{\ }particle system.
Through careful and detailed analytical derivation in suppletmental
material, we can get

\begin{eqnarray}
\psi _{1} &=&\frac{\sqrt{2}}{2}e^{-i\chi }\left( \psi _{l_{1}}\cos \kappa
t+i\psi _{l_{2}}\sin \kappa t\right) ,  \notag \\
\psi _{2} &=&\frac{\sqrt{2}}{2}e^{-i\chi }\left( i\psi _{l_{1}}\sin \kappa
t+\psi _{l_{2}}\cos \kappa t\right) ,
\end{eqnarray}

and
\begin{equation}
{\psi }_{\tau }=\frac{\sqrt{2}}{2}e^{-i\chi }\left( \psi _{l_{1}}+\psi
_{l_{2}}\right) e^{i\kappa t}.
\end{equation}

The specific expression of $\chi $ is in the supplemental material. The
initial wave functions are $\psi _{1}(0)=\frac{\sqrt{2}}{2}\psi _{l_{1}}(0)$%
, $\psi _{2}(0)=\frac{\sqrt{2}}{2}\psi _{l_{2}}(0)$ and ${\psi }_{\tau }{(0)}%
=\frac{\sqrt{2}}{2}[{\psi _{l_{1}}(0)+\psi _{l_{2}}(0)]}$ of particles with
CVSS in Eqs. (2) and (3). This indicates $\psi _{l_{1}}$\ and $\psi _{l_{2}}$%
\ are been coupled to form $\psi _{1}$\ and $\psi _{2}$\ linearly. It is
shown that the linear superposition between quantum states is not only a
mathematical operation, but also corresponds to a real physical process. The
difference between ${\psi }$ and ${\psi }_{\tau }$ is the latter shows a
formation process of the superposition state,\textbf{\ }and LVSS is the
initial state of CVSS. A measurement of particles breaks the coupling
between quantum states, resulting in $\kappa \rightarrow 0$, then $\psi
_{1}\rightarrow \frac{\sqrt{2}}{2}e^{-i\chi }\psi _{l_{1}}$ , $\psi
_{2}\rightarrow \frac{\sqrt{2}}{2}e^{-i\chi }\psi _{l_{2}}$ and CVSS decays
to LVSS\textbf{. }This is what the Copenhagen School says that a measurement
will causes collapse of a superposition state. This idea is also true for
CVSS with different proportional coefficients of $\psi _{1}$ and $\psi _{2}$
which is described in supplemental material. Note that the real motion of
particles should be described by $\psi _{1}$ or $\psi _{2}$. Moreover,
numerical solutions have been taken with $\frac{\sqrt{2}}{2}\psi _{l_{1}}$
and $\frac{\sqrt{2}}{2}\psi _{l_{2}}$ as initial wave functions in Eq. (1)
and shown in\textbf{\ }Fig. S1 in supplemental material.

\textbf{Degeneracy of angular momentum in vortex---}We study the dynamics
evolution of a particle system with CVSS\emph{\ }with $l_{1}=l,l_{2}=0$
numerically and give the diagrams of $\left\vert {\psi }_{\tau }\right\vert
^{2}$ as shown in Fig. 2. We can see that there are $l$ quantum vortices in $%
\left\vert {\psi }_{\tau }\right\vert ^{2}$ profiles, but such quantum
vortex lattice were always generated in rotating BECs \cite{Vortex lattice
formation,Measuring the disorder,Exotic vortex lattices,PT symmetric gain
and loss,Stripes and honeycomb,Gauge potential induced,Vortex lattice in a
uniform,Rotating binary Bose Einstein,Vortex Lattice Formation in Dipolar},
this means that we find a new method for preparing quantum vortex lattice of
BEC. In order to understand the physical mechanism of the quantum vortex
lattice produced by this way, we investigate the dynamics evolution of a
particle system with a quantum vortex state ${\psi _{l}}$ numerically which
satisties $i{\frac{\partial \psi _{l}}{\partial t}}=[-\bigtriangledown
_{\perp }^{2}+V(r)]\psi _{l}$, and we have given the numerical solution
diagrams of $\left\vert {\psi }_{l}\right\vert ^{2}$ and $\arg (\psi _{l})$
in Fig. 2.

\begin{figure}[th]
\centering\includegraphics[width=6cm,height=4cm]{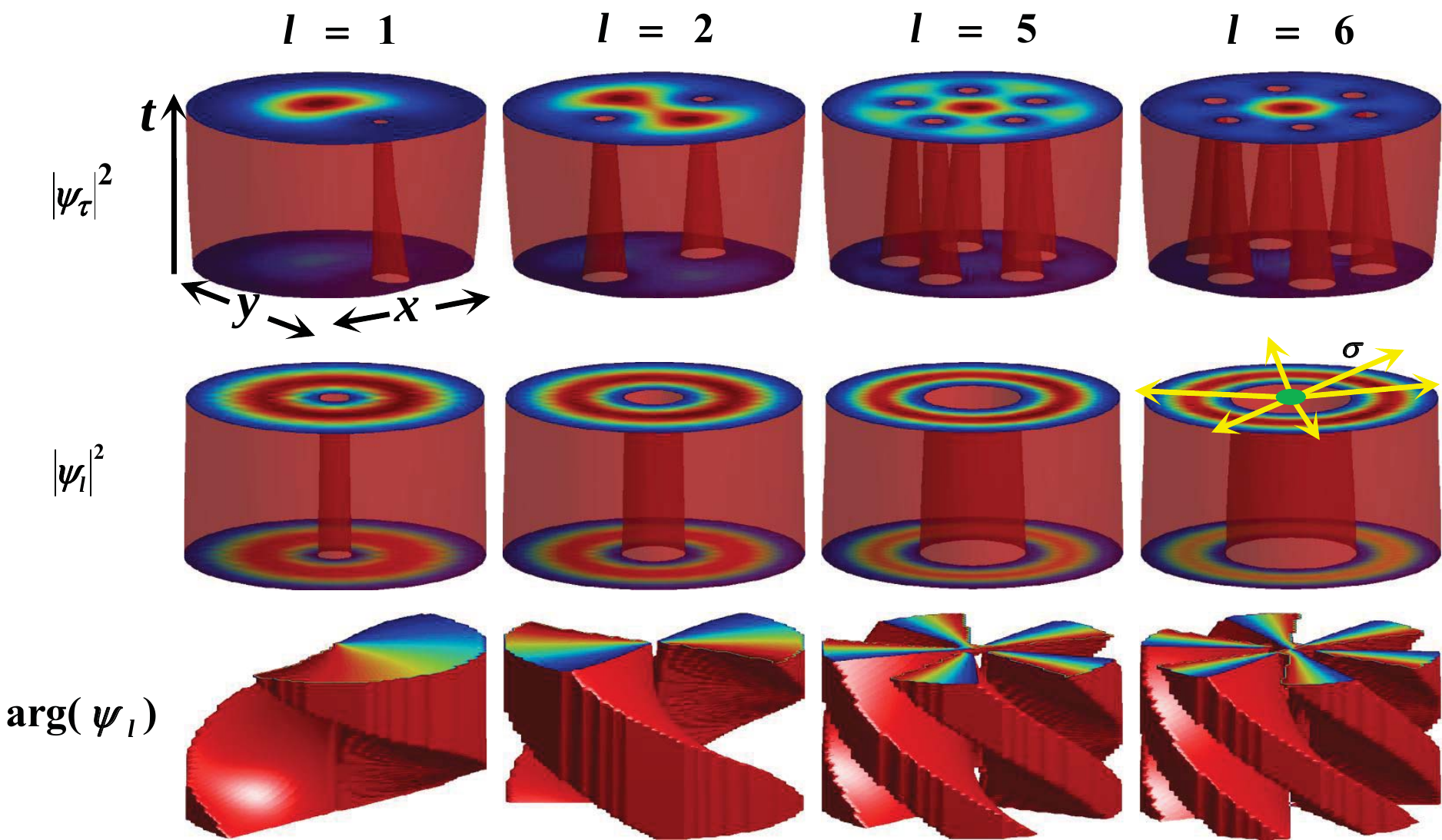}
\caption{The dynamic evolution diagrams of particle system with CVSS or
single vortex state. The parameters of $\left\vert \protect\psi _{\protect%
\tau }\right\vert ^{2}$ are $l_{1}=l,l_{2}=0,g_{1}=g_{2}=0,g_{12}=0,\protect%
\sigma _{1}=\protect\sigma _{2}=2,\protect\kappa =25\protect\pi .$ The
parameter of $\left\vert \protect\psi _{l}\right\vert ^{2}$ and $\arg \left(
\protect\psi _{l}\right) $ is $\protect\sigma =2.$ $V_{0}=0.5$ holds for
both cases. The green point in $\left\vert \protect\psi _{l}\right\vert ^{2}$
profile with $l=6$ denotes a particle source, the yellow arrows indicate
that the particle source emits particles, and $\protect\sigma $ is the size
of the scale that the particles can be reach in space.}
\end{figure}

$\left\vert {\psi }_{l}\right\vert ^{2}$ profiles exhibit a single vortex
which has been realized experimentally \cite{Vortices in a Bose-Einstein
Condensate}, and phase $\arg (\psi _{l})$ distributions appear $l$ vortices,
which prompts us to establish a new idea, that is to do the transformation $%
e^{\pm il\theta }\rightarrow \left( e^{\pm i\theta }\right) ^{l}.$\textbf{\ }%
$e^{\pm il\theta }$ characterizes the real space perimeter of ${\psi _{\pm l}%
}$ state could accommodate an integer number $l$ of de Broglie wavelengths
of a particle. However, $\left( e^{\pm i\theta }\right) ^{l}$ is refer to
the real space perimeter of ${\psi _{\pm l}}$ state could accommodate an
integer number $l$ of particles which moving in a circular motion, or a
particle has $l$ degeneracy spatial positions, and $l=0$ can be regarded as
a vacuum state. $e^{\pm il\theta }\rightarrow \left( e^{\pm i\theta }\right)
^{l}$ also cause $\pm l\hbar \rightarrow \Sigma _{1}^{l}\left( \pm \hbar
\right) $, the significance of this operation is that we regard $\pm l\hbar $
as an angular momentum level, the particles at this level have angular
momentum $\pm \hbar $ and\ $l$\ degree degenerate spatial positions, we call
this is degeneracy of angular momentum in vortex (DAMV).\ A particle with $%
\psi _{l}$ state makes a circular motion in real space with angular momentum
$\hbar $ which named intrinsic angular momentum (IAM), and by $l$ degenerate
spatial positions on this circle for the particle. So $\left\vert {\psi }%
_{l}\right\vert ^{2}$ profiles display a single vortex and phase $\arg (\psi
_{l})$ distributions shown $l$ vortices. IAM is relative to the spatial
center of the particle system with CVSS, so it has the characteristic of
\textquotedblleft spin without ratate\textquotedblright\ relative to a
particle itself. We think that the continuous movement of the mass element
in the BEC is the core to produce a quantum vortex lattice. When the mass
element has an angular momentum $\hbar $ relative to the center of the
condensate a small vortex is formed, so the number of vortex core in the
vortex lattice is $l$.

Now we study the actual movement of particles with CVSS, according to the
average momentum formula of a particle with $\psi _{1}$ or $\psi _{2}$ state
in Eq. (2), we can get
\begin{eqnarray}
L_{1z} &=&\frac{1}{2}[l_{1}|c_{1,l_{1}}(t)|^{2}+l_{2}|c_{1,l_{2}}(t)|^{2}]%
\hbar  \notag \\
&=&l_{1}\frac{\hbar }{2}-\left( l_{1}-l_{2}\right) \frac{\hbar }{2}\sin
^{2}\kappa t,  \notag \\
L_{2z} &=&\frac{1}{2}[l_{1}|c_{2,l_{1}}(t)|^{2}+l_{2}|c_{2,l_{2}}(t)|^{2}]%
\hbar  \notag \\
&=&l_{1}\frac{\hbar }{2}-\left( l_{1}-l_{2}\right) \frac{\hbar }{2}\cos
^{2}\kappa t.
\end{eqnarray}

1) As $l_{1}=l,l_{2}=0,$ then $L_{1z}=l\left( \frac{\hbar }{2}\cos
^{2}\kappa t\right) =\Sigma _{1}^{l}\left( \frac{\hbar }{2}\cos ^{2}\kappa
t\right) $, $L_{2z}=l\left( \frac{\hbar }{2}\sin ^{2}\kappa t\right) =\Sigma
_{1}^{l}\left( \frac{\hbar }{2}\sin ^{2}\kappa t\right) $. This shows that
the particles which in the quantum vortex\ lattice have an IAM of $\frac{%
\hbar }{2}\cos ^{2}\kappa t$ or $\frac{\hbar }{2}\sin ^{2}\kappa t$\textbf{.
}Now we consider the effect of the interaction which is described by $\kappa
\rightarrow 0$ between any one finite ($\sigma \neq \infty $), closed
particle system and the vacuum ($l=0$ , $\sigma =\infty $). According to Eq.
(S7) (See the supplemental material for specific derivation) we can be obtain

\begin{equation}
\psi _{\tau }=e^{-i\frac{1}{2}(\alpha +g\beta -\mu )t}\psi _{\pm l}.
\end{equation}%
We can see from Eq. (5) that due to the coupling between the system and the
vacuum, there is an additional phase in its wave function, and this phase
indicates that the system has a vortex phase space \cite{Holonomy the
Quantum Adiabatic Theorem and Berrys Phase}, here $\alpha =\frac{|l|+1}{%
\sigma ^{2}}$.

2) As $l_{1}=-l_{2}=l,$ then $L_{1z}=l\left( \frac{\hbar }{2}\cos 2\kappa
t\right) =\Sigma _{1}^{l}\left( \frac{\hbar }{2}\cos 2\kappa t\right) $, $%
L_{2z}=l\left( -\frac{\hbar }{2}\cos 2\kappa t\right) =\Sigma _{1}^{l}(-%
\frac{\hbar }{2}\cos 2\kappa t)$. This shows that the particle with CVSS
have an IAM of $\frac{\hbar }{2}\cos 2\kappa t$ or $-\frac{\hbar }{2}\cos
2\kappa t$, this just illustrates the characteristics of $\arg (\psi _{1})$
and $\arg (\psi _{2})$ distributions as shown in Fig. S1 in suppletmental
material, they also show that there are coupled particle pairs with IAM of $%
\frac{\hbar }{2}\cos 2\kappa t$ and $-\frac{\hbar }{2}\cos 2\kappa t$ in
multi-particle system with CVSS\textbf{.} The formation and existence of
these pairs are not limited by the spatial location and the influence of
scattering interaction between particles. According to Eq. (1) and its
initial conditions, we know that this is a method for preparing entangled
state \cite{Quantum squeezing and entanglement in a two-mode Bose-Einstein
condensate with}. This means that we find a hidden variable $\kappa $\ in
the entanglement of particles, which corresponds to a linear interaction.\
In addition, due to the distance between entangled particles is not
infinite, we can consider this entangled system to be a finite and closed
system, so that the non-local entangled system is localized into a finite
vortex space.\textbf{\ }See the supplemental material for detail.

\begin{figure}[th]
\centering\includegraphics[width=6cm,height=2cm]{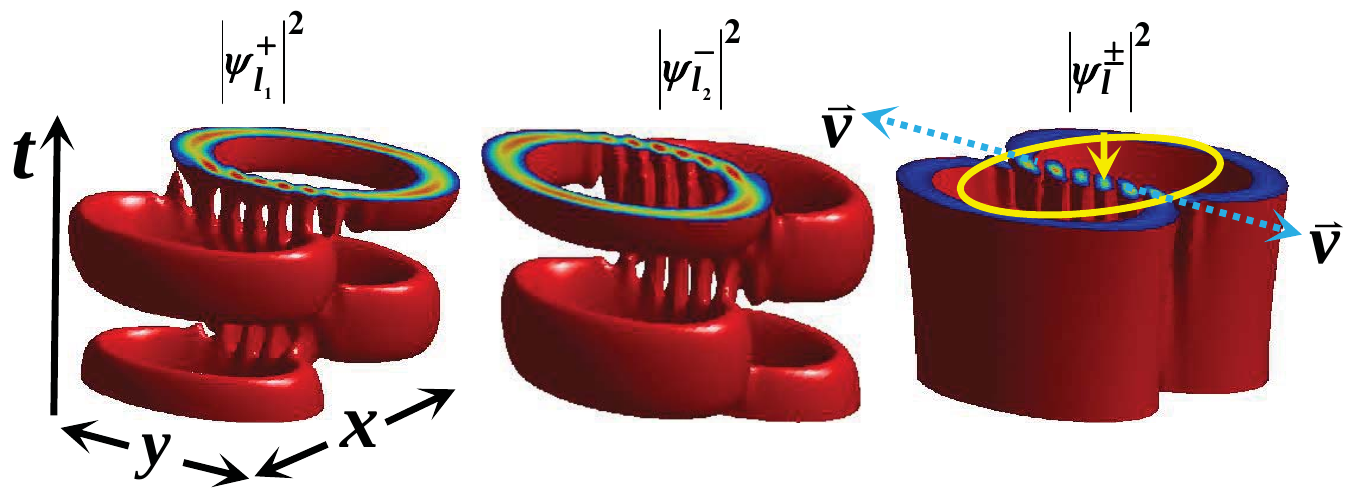}
\caption{Numerical simulation of a free particle with $l_{1}=l_{2}=13,$ $%
r_{0}=r_{0}(2,0),$ $\protect\sigma _{1}=\protect\sigma _{2}=0.5,\protect%
\kappa =15\protect\pi ,$ for different $\protect\varpi =0,3,-3$, there are
the same diagrams. The yellow circle in $\left\vert \protect\psi _{l}^{\pm
}\right\vert ^{2}$ profile is regard the particle density distribution in
space as a whole, the yellow arrow points to the particle's spatial
position, and the blue arrows show the possible direction of the movement of
the particle.}
\end{figure}

\textbf{Free particle and double-slit interference experiment}---We now
consider another special case, that is, $l_{1}=l_{2}=\pm l,$ then\textbf{\ }$%
L_{1z}=L_{2z}=\Sigma _{1}^{l}(\pm \hbar /2)$. This situation reflects the
continuous movement which is described by $\psi _{1}=\psi _{2}=\frac{\sqrt{2}%
}{2}e^{i\varpi t}e^{-i(\alpha -\mu )t}e^{i\kappa t}\psi _{\pm l}$ of a
particle on the circumference of a eigen vortex $\psi _{\pm l}$, $\varpi $
denotes the circular frequency of the particle. In fact, here we close the
external potential well and take the scattering interaction between
particles to be zero, this means that our research object is a free
particle. In addition, from the perspective of classical mechanics, a
particle moving in a circular motion must be subjected to a centripetal
force. We suggest that the provider is the continuous and symmetrical nature
of space for micro particle, according to this idea, it can be known that
the IAM of a particle has nothing to do with the direction of measurement,
and its root cause is that the measurement surface has symmetric property
about the direction of motion for a particle\textbf{. }Note that the IAM
value of a free particle is determined and will not change due to
measurement. And the IAM of $\hbar /2$ or $-\hbar /2$ of a particle is
independent on the moving direction that described by $\varpi $ and even
whether or not moving in space. In order to express this idea intuitively,
we have designed a scheme and given relevant numerical solution diagrams as
shown in Fig. 3. The specific analysis diagram Fig. S3 and numerical
solution scheme are in the supplemental material. Furthermore, considering $%
\left\vert {\psi }_{l}^{\pm }\right\vert ^{2}$ profile as a whole, the free
particle motion can be understood as the motion of the mass center of a
ring-shaped matter, and its wave function is a vortex function as shown in
Fig. S3(b). This may be the reason for the A-B effect \cite{Significance of
electromagnetic potentials in the quantum theory} in the double-slit
interference experiments of particles.

Our analysis of free particles is very novel, so it is necessary to verify
its correctness. The best way is to carry out double-slit interference (DSI)
experiment of \textquotedblleft free particles\textquotedblright , various
versions of which-way experiment show that as long as the path information
of particles is obtained, the interference fringes on screen will disappear
\cite{Delayed Choice Quantum Eraser,Delayed choice gedanken,Observing
momentum disturbance,Which way double slit,Wave particle duality}. Now we
give a new explanation about this phenomenon. We think the distribution
pattern of particles on the screen is dependent on the nature of the
particle source which is in front of screen, but the slits are the paths
that the particles reach the screen. We consider the case of $l_{1}=-l_{2}=l$
in CVSS and extend $\sigma _{1}=\sigma _{2}=\sigma $, $\sigma $ denotes the
size of the spatial scale that particles can be reach after they are emitted
from the particle source. We think that the particles emitted from the
particle source have an IAM of $\frac{\hbar }{2}\cos 2\kappa t$ or $-\frac{%
\hbar }{2}\cos 2\kappa t$, so the distribution of particles in space meet
the distribution rules similar to $\left\vert \psi _{1}\right\vert
^{2}\oplus \left\vert \psi _{2}\right\vert ^{2}$ and $\left\vert \psi _{\tau
}\right\vert ^{2}$ profiles as shown in Fig. S1 in supplemental material.
When particles pass through the left or right slit to the screen, the
distribution pattern is interference fringes. Here $\left\vert \psi
_{1}\right\vert ^{2}\oplus \left\vert \psi _{2}\right\vert ^{2}$
characterizes entangled two-particles interference but not two particle
interference \cite{Can Two-Photon Interference be Considered the
Interference of Two Photons?}, and $\left\vert \psi _{\tau }\right\vert ^{2}$
is one-particle interference. The meaning of taking the straight sum is that
the two particles in the entangled state have their own spaces. When the
particles are measured, it results in $\kappa \rightarrow 0$ and $\psi
_{1}\rightarrow \frac{\sqrt{2}}{2}e^{-i\chi }\psi _{l_{1}}$, $\psi
_{2}\rightarrow \frac{\sqrt{2}}{2}e^{-i\chi }\psi _{l_{2}}$, which in turn
produce new particle sources $\psi _{l_{1}}$ and $\psi _{l_{2}}$. Combined
with the analysis of free particles, the particles emitted from this
particle source have an IAM of $\hbar /2$ or $-\hbar /2$, and the
distribution of particles in space is similar to the pattern which marked
with yellow arrows as shown in Fig. 2. Therefore, quantum measurement is
performed on the particles before or after the double slit that named QMBDSI
and QMADSI respectively, the interference fringes on the screen will
disappear. We find the dynamic law of \textquotedblleft free
particles\textquotedblright\ which is different from the guided wave theory
\cite{Experimental nonlocal and surreal,Observing the Average
Trajectories,Interfering Quantum Trajectories,Experimental nonlocal steering}%
\textbf{.} We have given numerical simulation diagrams of one-particle and
entangled two-particles dynamic evolution in the DSI, QMBDSI and QMADSI as
shown in Fig. 4. The specific analysis diagram Fig. S4 and\textbf{\ }scheme
of numerical simulation is described in detail in the supplemental material.

$\left\vert {\psi }_{_{l_{1}}}^{L}\right\vert ^{2}$ and $\left\vert {\psi }%
_{_{l_{2}}}^{R}\right\vert ^{2}$ profiles at moment $t=10,40$ indicate that
the particles pass through the left or right slit individually. Comparing
the profiles of $\left\vert {\psi }_{_{l_{1}}}^{L}+{\psi }%
_{_{l_{2}}}^{R}\right\vert ^{2}$ and $\left\vert {\psi }_{_{l_{1}}}^{L}%
\right\vert ^{2}\oplus \left\vert {\psi }_{_{l_{2}}}^{R}\right\vert ^{2}$ in
the DSI at $t=15,35$, we know that the line width of the interference fringe
of entangled two-particles is half of that of one-particle, this rule has
been confirmed in the photon interference experiment \cite{Measurement of
the Photonic de Broglie Wave length of Entangled Photon}. Both $\left\vert {%
\psi }_{_{l_{1}}}^{L}+{\psi }_{_{-l_{2}}}^{R}\right\vert ^{2}$ and $%
\left\vert {\psi }_{_{l_{1}}}^{L}\right\vert ^{2}\oplus \left\vert {\psi }%
_{_{-l_{2}}}^{R}\right\vert ^{2}$ profiles in the QMBDSI exhibit a normal
distribution, while $\left\vert {\psi }_{_{l_{1}}}^{L}\right\vert ^{2}\oplus
\left\vert {\psi }_{_{l_{2}}}^{R}\right\vert ^{2}$ profiles in the QMADSI
show a distribution pattern which corresponding to double slits. The
numerical simulation diagrams of particle interference in DSI, QMBDSI, and
QMADSI are consistent with our analysis of the double-slit interference
experiment\textbf{. }This means that the IAM of particles determine whether
interference fringes appear on the screen. In addition, when we consider a
free particle as the mass center of a ring-shaped matter, the double-slit
interference becomes the interference of two eigen vortices. The double
vortex interference shows the same law as the double-slit interference, and
points out the existence of single-slit diffraction of particles. The
specific implementation scheme and numerical simulation diagrams are in the
supplemental material.

\begin{figure}[th]
\centering\includegraphics[width=7cm,height=7.5cm]{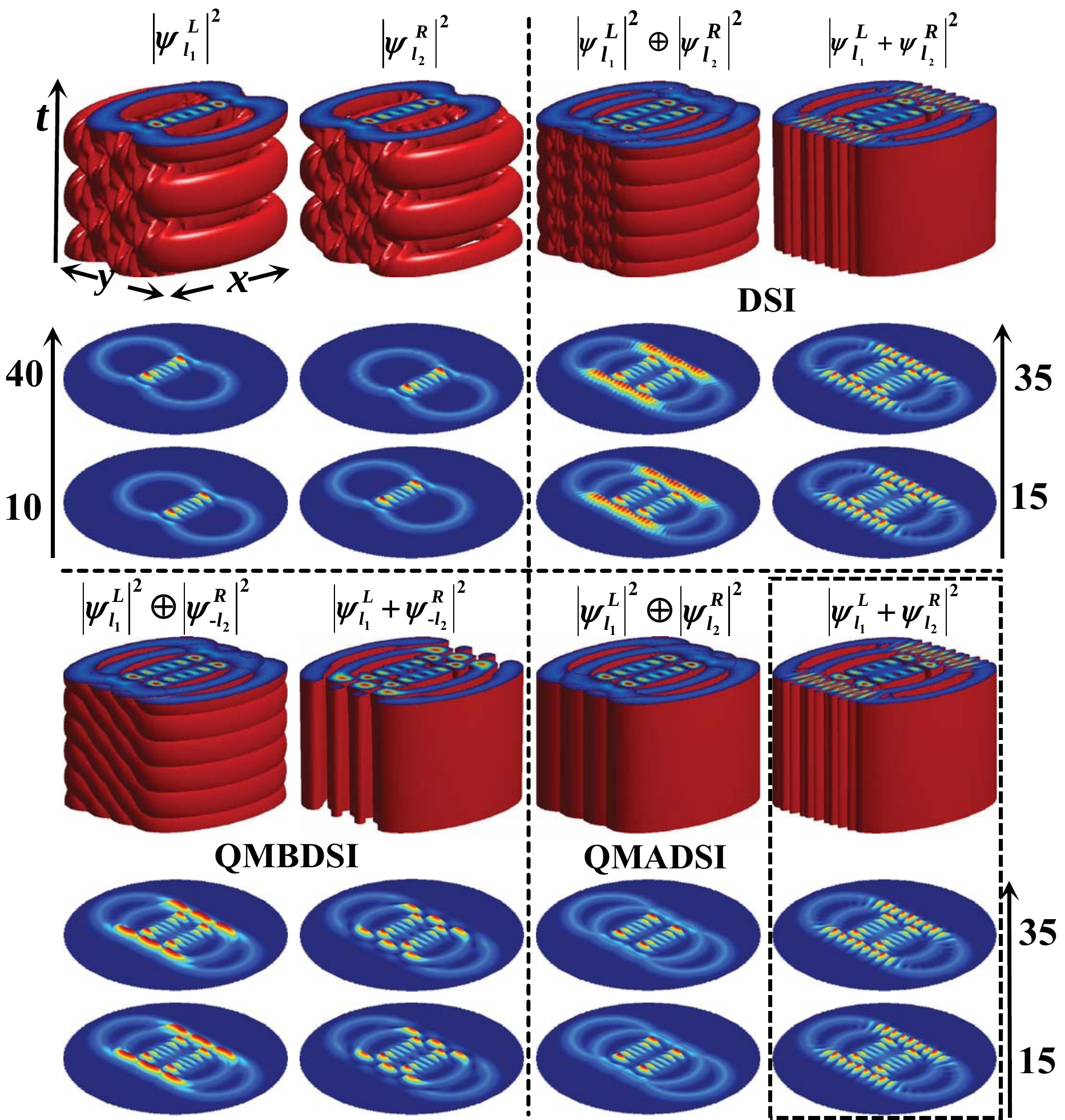}
\caption{Numerical simulation of density distribution in double-slit
interference experiment of entangled two-particles and one-particle which
described by $\left\vert \protect\psi _{l_{1}}^{L}\right\vert ^{2}\oplus
\left\vert \protect\psi _{l_{2}}^{R}\right\vert ^{2}$ and $\left\vert
\protect\psi _{l_{1}}^{L}+\protect\psi _{l_{2}}^{R}\right\vert ^{2}$
respectively with $r_{1}=r_{1}(0,3),$ $r_{2}=r_{2}(0,9),$ $l_{1}=-l_{2}=13,$
$\protect\sigma =2,$ $\protect\kappa _{0}=50\protect\pi . $ $\protect\kappa %
=50\protect\pi $ for DSI and QMBDSI, and $\protect\kappa =0.05\protect\pi $
for QMADSI. $\left\vert \protect\psi _{l_{1}}^{L}+\protect\psi %
_{l_{2}}^{R}\right\vert ^{2}$ profiles which in the dashed box is not
allowed physically, (Due to $\protect\kappa \rightarrow 0$, the particles
can only be with $\protect\psi _{l_{1}}^{L}$ or $\protect\psi _{l_{2}}^{R}$
state.) although the interference fringe distribution can be obtained in
math.}
\end{figure}

\textbf{Conclusion---}This letter have studed the analytical solution of
CVSS in detail, and we find that there is a formation process for
superposition state, not just the algebraic sum of states. The essence of
superposition state collapse is the coupling coefficient $\kappa \rightarrow
0$ in CVSS caused by quantum measurement. Secondly, we propose DAMV to
clarify the fundamental principle of the appearance of vortex lattice in
BEC, and localized the non-local quantum entanglement system into a vortex
space. In addition, we suggest that a free particle have a vortex wave
function, which is verified by the numerical simulation of a double-slit and
double vortex interference experiment of particles.

This work was supported by NSFC under grants Nos.12074027.


\begin{thebibliography}{99}
\bibitem{An Undulatory Theory of the Mechanics of Atoms and Molecules} E.
Schr\"{o}dinger, Phys. Rev. \textbf{28}, 1049 (1926).

\bibitem{Quantum entanglement} R. Horodecki, P. Horodecki, M. Horodecki, K.
Horodecki, Rev. Mod. Phys. \textbf{81}, 865 (2009).

\bibitem{18 Qubit Entanglement} X. L. Wang, Y. H. Luo, H. L. Huang, M. C.
Chen, Z. E. Su, C. Liu, C. Chen, W. Li, Y. Q. Fang, X. Jiang, J. Zhang, L.
Li, N. L. Liu, C. Y. Lu, and J. W. Pan, Phys. Rev. Lett. \textbf{120},
260502 (2018).

\bibitem{Experimental quantum teleportation} D. Bouwmeester, J. W. Pan, K.
Mattle, M. Eibl, H. Weinfurter and A. Zeilinger, Nature \textbf{390}, 575
(1997).

\bibitem{Entanglement of photons} A. S. Rab, E. Polino, Z. X. Man, N. B. An,
Y. J. Xia, N. Spagnolo, R. L. Franco and F. Sciarrino, Nature \textbf{8},
915 (2017).

\bibitem{Einstein Podolsky Rosen} J. C. Lee, K. K. Park, T. M. Zhao and Y.
H. Kim, Phys. Rev. Lett. \textbf{117}, 250501 (2016).

\bibitem{Entanglement of photons with complex} J. Tang, Y. Ming, Z. X. Chen,
W. Hu, F. Xu and Y. Q. Lu, Phys. Rev. A \textbf{94}, 012313 (2016).

\bibitem{Can Quantum Mechanical Description} A. Einstein, B. Podolsky, N.
Rosen, Phys. Rev. \textbf{47}, 777 (1935).

\bibitem{A Suggested Interpretation of the Quantum Theory in Terms of Hidden Variables I and II} %
D. Bohm, Phys. Rev. \textbf{85}, 166 (1952).

\bibitem{Structure and generation of the vortex-antivortex superposed state in Bose-Einstein condensates} %
M. Liu, L. H. Wen, H. W. Xiong and M. S. Zhan, Phys. Rev. A \textbf{73},
063620 (2006).

\bibitem{The Vortex Phase Qubit} K. T. Kapale and J. P. Dowling, Phys. Rev.
Lett. \textbf{95}, 173601 (2005).

\bibitem{Structure of two-component Bose-Einstein condensates with respective} %
L. H. Wen, Y. J. Qiao, Y. Xu and L. Mao, Phys. Rev. A \textbf{87}, 033604
(2013).

\bibitem{Optical control of the internal and external angular momentum of a Bose-Einstein condensate} %
K. C. Wright, L. S. Leslie, and N. P. Bigelow, Phys. Rev. A \textbf{77},
041601(R) (2008).

\bibitem{Quantized Rotation of Atoms From Photons with Orbital Angular Momentum} %
M. F. Andersen, C. Ryu, P. Clad\'{e}, V. Natarajan, A.Vaziri, K. Helmerson,
and W. D. Phillips, Phys. Rev. Lett. \textbf{97}, 170406 (2006).

\bibitem{Sculpting the Vortex State of a Spinor BEC} K. C. Wright, L. S.
Leslie, A. Hansen, and N. P. Bigelow, Phys. Rev. Lett. \textbf{102}, 030405
(2009).

\bibitem{Spontaneous breaking of spatial and spin symmetry in spinor condensates} %
M. Scherer, B. L\"{u}cke, G. Gebreyesus, O. Topic, F. Deuretzbacher, W.
Ertmer, L. Santos, J. J. Arlt, and C. Klempt, Phys. Rev. Lett. \textbf{105},
135302 (2010).

\bibitem{Arbitrary Coherent Superpositions of Quantized Vortices} S.
Thanvanthri, K. T. Kapale and J. P. Dowling, Phys. Rev. A \textbf{77},
053825 (2008).

\bibitem{Vortex lattice formation} M. Tsubota, K. Kasamatsu, M. Ueda, Phys.
Rev. A \textbf{65}, 023603 (2002).

\bibitem{Measuring the disorder} A. Rakonjac, A. L. Marchant, T. P. Billam,
J. L. Helm, M. M. H. Yu, S. A. Gardiner, and S. L. Cornish\textit{,} Phys.
Rev. A \textbf{93}, 013607 (2016).

\bibitem{Exotic vortex lattices} X. F. Zhang, L. Wen, C. Q. Dai, R. F. Dong,
H. F. Jiang, H. Chang and S. G. Zhang, Sci. Rep.\textbf{\ 6, }19380, (2016).

\bibitem{PT symmetric gain and loss} D. Haag, D. Dast, H. Cartarius and G.
Wunner, Phys. Rev. A \textbf{97}, 033607 (2018).

\bibitem{Stripes and honeycomb} K. Kasamatsu and K. Sakashita, Phys. Rev. A
\textbf{97}, 053622 (2018).

\bibitem{Gauge potential induced} J. J. Jin, W. Han and S. Y. Zhang, Phys.
Rev. A \textbf{98}, 063607 (2018).

\bibitem{Vortex lattice in a uniform} S. K. Adhikari, Phys.: Condens.
Matter. \textbf{31,} 275401 (2019).

\bibitem{Rotating binary Bose Einstein} M. N. Tengstrand, P. Sturmer, E. O.
Karabulut and S. M. Reimann, Phys. Rev. Lett. \textbf{123}, 160405 (2019).

\bibitem{Vortex Lattice Formation in Dipolar} S. B. Prasad, T. Bland, B. C.
Mulkerin, N. G. Parker and A. M. Martin, Phys. Rev. A \textbf{100}, 023625
(2019).

\bibitem{Vortices in a Bose-Einstein Condensate} M. R. Matthews, B. P.
Anderson, P. C. Haljan, D. S. Hall, C. E. Wieman, and E. A. Cornell, Phys.
Rev. Lett. \textbf{83}, 2498 (1999).

\bibitem{Holonomy the Quantum Adiabatic Theorem and Berrys Phase} B. Simon,
Phys. Rev. Lett.\textbf{\ 51}, 2167 (1983).

\bibitem{Quantum squeezing and entanglement in a two-mode Bose-Einstein condensate with} %
S. Choi and N. P. Bigelow, Phys. Rev. A \textbf{72}, 033612 (2005).

\bibitem{Significance of electromagnetic potentials in the quantum theory} %
Y. Aharonov, D Bohm, Phys. Rev. \textbf{115}, 485, (1959).

\bibitem{Delayed Choice Quantum Eraser} Y. H. Kim, R. Yu, S. P. Kulik, Y.
Shih and M. O. Scully, Phys. Rev. Lett. \textbf{84, }1\textbf{\ }(2000).

\bibitem{Delayed choice gedanken} X. S. Ma, J. Kofler, and A. Zeilinger,%
\textit{\ }Rev. Mod. Phys. \textbf{88}, 015005 (2016).

\bibitem{Observing momentum disturbance} Y. Xiao, H. M. Wiseman, J. S Xu, Y.
Kedem, C. F. Li, and G. C. Guo,\textit{\ }Sci. Adv. \textbf{5}, eaav9547
(2019).

\bibitem{Which way double slit} J. Q. Quach,Phys. Rev. A \textbf{95}, 042129
(2017).

\bibitem{Wave particle duality} E. Bagan, J. A. Bergou and M. Hillery, Phys.
Rev. A \textbf{102}, 022224 (2020).

\bibitem{Can Two-Photon Interference be Considered the Interference of Two Photons?} %
T. B. Pittman, D. V. Strekalov, A. Migdall, M. H. Rubin, A. V. Sergienko and
Y. H. Shih, Phys. Rev. Lett. \textbf{77, }1917\textbf{\ }(1996).

\bibitem{Experimental nonlocal and surreal} D. H. Mahler, L. Rozema, and K.
Fisher, L. Vermeyden, K. J. Resch, H. M. Wiseman and A. Steinberg,\textit{\ }%
Sci. Adv. \textbf{2,}\ e1501466 (2016).

\bibitem{Observing the Average Trajectories} S. Kocsis, B. Braverman, S.
Ravets, M. J. Stevens, R. P. Mirin, L. K. Shalm and A. M. Steinberg,\textit{%
\ }Science \textbf{332, }1170 (2011).

\bibitem{Interfering Quantum Trajectories} K. Mathew and M. V. John,
Foundations of Physics \textbf{47}, 873 (2017).

\bibitem{Experimental nonlocal steering} Y. Xiao, Y. Kedem, J. S. Xu, C. F.
Li and G. C. Guo, Optics Express \textbf{25}, 14463 (2017).

\bibitem{Measurement of the Photonic de Broglie Wave length of Entangled Photon} %
K Edamatsu, R Shimizu, T Itoh, Phys. Rev. Lett. \textbf{89}, 213601 (2002).
\end{thebibliography}
\end{document}